\documentstyle[prb,aps]{revtex}
\begin{document}
\twocolumn[
\hsize\textwidth\columnwidth\hsize\csname %
@twocolumnfalse\endcsname

\draft
\widetext
\title{
Observation of the antiferroquadrupolar order in DyB$_2$C$_2$ 
by resonant x-ray scattering
}
\author{T. Matsumura, N. Oumi and K. Hirota}
\address{Department of Physics, Tohoku University, Sendai, 980-8578, Japan}
\author{H. Nakao and Y. Murakami\cite{adMura}}
\address{Photon Factory, Institute of Materials Structure Science, High Energy Accelerator Research Organization, 
Tsukuba, 305-0801, Japan}
\author{Y. Wakabayashi\cite{adWaka}}
\address{Department of Physics, Faculty of Science and Technology, Keio University, 
3-14-1 Hiyoshi, Yokohama, 223-8522, Japan}
\author{T. Arima}
\address{Institute of Materials Science, University of Tsukuba, Tsukuba, 305-8573, Japan}
\author{S. Ishihara\cite{adIshi} and Y. Endoh}
\address{Institute for Materials Research, Tohoku University, Sendai, 980-8577, Japan}
\date{August 28, 2001}
\maketitle
\begin{abstract}
We have investigated the 
antiferroquadrupolar(AFQ) order in DyB$_2$C$_2$ by resonant x-ray scattering. 
X-rays with energies near the $L_3$ absorption edge of Dy were employed. 
Superlattice peaks that correspond to three kinds of propagation vectors of (1 0 0), (1 0 1/2) 
and (0 0 1/2) were investigated in detail with polarization analyses. The experimental results 
are analyzed using a formalism on resonant x-ray scattering and a model of the AFQ 
order. The magnetic and quadrupolar scatterings are explained by the model satisfactorily. 
A detailed investigation on the critical behavior of the AFQ ordering is also reported. The critical 
exponent $\beta$ is deduced to be 0.35, not far from the three-dimensional Heisenberg system. 
We have also succeeded in detecting the diffuse scattering above $T_{Q}$. 
\end{abstract}

\phantom{.}
]
\narrowtext

% body of paper here

\section{Introduction}
There is a class of magnetic materials in which the orbital degree of freedom remains 
due to highly symmetrical crystalline electric fields (CEF). Many of such materials 
undergo phase transitions resulting in the lifting of the orbital degeneracy. 
In $f$-electron systems, when the CEF ground state has a non-Kramers-type degeneracy, 
there is a possibility that a periodic ordering of the anisotropic charge distributions of 
the $f$-electrons takes place, other than magnetic orderings. This asphericity 
can be represented in the lowest order by the quadrupolar moment of the localized 
$f$-electrons. The orderings are driven by quadrupolar pair interactions in combination 
with magneto-elastic 
interactions.~\cite{Levy} Even in materials in which magnetic interactions dominate, 
there are many cases where the magnetic properties are largely influenced by 
underlying quadrupolar interactions.~\cite{Morin} 
To study the mechanism of these phenomena the observation of quadrupolar moments, 
with changing sample environments such as temperature, pressure, and magnetic field 
is necessary. 

In order to observe the asphericity of charge distribution of an ion, diffraction 
method is essential, especially for antiferroquadrupolar (AFQ) orders which 
normally do not accompany measurable uniform lattice distortions. 
Various methods using x-rays and neutrons have been studied. The most direct way 
to observe the asphericity of an electron shell is to use x-ray Thomson scattering in 
non-resonant region. Keating's experiment on holmium succeeded in detecting the scattering of x-rays 
from the spiral arrangement of aspherical $4f$-shells occurring at twice the 
magnetic propagation vector.~\cite{Keating} The same method was formulated 
by Amara and Morin using multipole expansion, giving a description of the 
scattering in terms of quadrupolar moments of ions.~\cite{Amara1} 
The AFQ order in NdMg was observed successfully.~\cite{Amara2} 
We also refer to a x-ray powder-diffraction combined with the maximum 
entropy method which succeeded in drawing a 
charge density map of a manganese oxide in an orbital ordered phase.~\cite{Takata} 

Neutron scattering can also be used to study the quadrupolar orders, 
although neutrons do not have a direct coupling with electric charges. 
Measurement of magnetic form factors can give information on 
aspherical charge distributions through spin densities. 
The first observation was performed by Ito and Akimitsu for K$_2$CuF$_4$.~\cite{Ito,Akimitsu} 
Felcher {\it et al.} measured the scattering-vector dependence and the temperature dependence 
of the aspherical part of the magnetic form factor, i.e., the $<j_2>$ term, of holmium 
by observing the third harmonic of the magnetic Bragg peak.~\cite{Felcher} 
In CeB$_6$ and TmTe, the existence of the AFQ phase with no magnetic order has been 
established through measurements of the induced antiferromagnetic (AFM) moments in 
magnetic fields.~\cite{Erkelens,Effantin,Link_1,Link_2,Mignot} Since the directions of the 
induced magnetic moments are coupled with the underlying quadrupolar moments, 
the quadrupolar order parameter can be estimated from symmetry arguments.~\cite{Shiina} 
However, both the form-factor measurement and the induced-moment measurement 
require magnetic field, which changes the unperturbed state in zero field. 

Appropriate method for each substance and purpose has been employed. In holmium, 
since it shows ideally the spiral magnetic structure, 
measurement of higher harmonics is directly connected to the aspherical component. 
Then the detailed measurement of the temperature dependence of the aspherical component 
by neutron scattering in zero magnetic field was possible.~\cite{Felcher} 
This is not the case for high symmetry systems like NdMg, CeB$_6$ and TmTe. 
We need to rely on x-ray scattering if we want to detect quadrupolar order in zero field, 
unless the sample exhibits periodic lattice distortion which is detectable with 
neutrons as in UPd$_3$.~\cite{McEwen} 
A great advantage of the non-resonant x-ray scattering is that we can obtain the absolute value 
of the ordered quadrupolar moment. However, the count rate could be so weak and it might be 
difficult to collect detailed data with changing sample environments.~\cite{Amara1,Amara2} 
In CeB$_6$ and TmTe, x-ray scattering experiments have been very difficult because of 
their low transition temperatures for x-ray experiments: $T_{Q}=$3.3 K for CeB$_6$ and 
1.8 K for TmTe. Further, no indication of periodic lattice distortion have been observed 
for the two compounds. Then the induced AFM structures in magnetic fields have been the 
clearest experimental evidence for the AFQ orders. 
Very recently, Nakao {\it et al.} succeeded in observing the AFQ order in CeB$_6$ 
by resonant x-ray scattering.~\cite{Nakao} 
Finally, form-factor measurements using x-rays and neutrons combined with 
the Fourier analysis or the maximum entropy method are quite effective when we 
want to investigate actual images of spin or charge distributions in real space. However, these are not 
suitable for the measurement of order parameters with changing sample environments.

The purpose of the present study is to observe the quadrupolar and magnetic order parameters 
of a rare-earth compound DyB$_2$C$_2$ in zero magnetic field. 
We have employed the resonant x-ray scattering for this purpose, 
which has recently been applied to the observations of the orbital orders in 
3$d$ transition metal oxides.~\cite{Murakami1,Murakami2} This method utilizes the 
characteristic that an atomic scattering factor is largely enhanced when the energy of an x-ray 
is tuned at an absorption edge of the atom. The first advantage of this method is the high 
count rate due to the enhancement at the absorption edge. This makes possible to measure 
temperature, azimuthal angle, polarization, and energy dependences of the peak intensities 
in detail.~\cite{Helgesen} Secondly, the element selectivity guarantees the resonant peak 
arise only from the element in study without doubt. Thirdly, we can distinguish periodic arrangement 
of the quadrupolar moments from periodic lattice distortions by measuring the energy dependence 
of the superlattice peak. Finally, the experiment can be performed with a small piece of single crystal 
using natural boron, which makes neutron scattering experiment difficult. 
An disadvantage is firstly that this method is not as direct as the non-resonant x-ray Thomson scattering; 
it is not possible to deduce the absolute value of the moment. Secondly, the scattering mechanism 
has not yet been solidly established. 
Several scattering mechanisms in the orbital ordered state are proposed from different viewpoints, 
particularly in 3$d$ compounds. ~\cite{Ishihara1,Ishihara2,Elfimov,Benfatto,Takahashi,Ishihara3} 
In the present study we simply assume that the $5d$ state is most influenced by the 
local $4f$ electrons through the Coulomb and exchange interactions, which 
we believe is the natural interpretation of the resonance in 
$4f$-electron systems. We will analyze 
the data quantitatively using the formulations developed by Blume.~\cite{Blume1,Blume2}

DyB$_2$C$_2$, with the tetragonal LaB$_2$C$_2$-type structure, 
is a compound that has recently been investigated in detail by Yamauchi {\it et al.} and is considered 
to show an AFQ order.~\cite{Yamauchi} 
Two phase transitions are clearly observed at 15~K and at 25~K with 
an entropy release of $R \ln 2$ and $R \ln 4$, respectively. 
The three phases are named phase I for $T>$25 K, phase II 
for 15 K$<T<$25 K, and phase III for $T<$15 K. 
No magnetic order appears in the phase II. 
The magnetic order in the phase III exhibits an unusual magnetic structure shown 
in Fig.~\ref{fg:structure}. 
The structure is represented by the four propagation vectors: 
$\bbox{k}_1=$(1 0 0), $\bbox{k}_2=$(1 0 1/2), $\bbox{k}_3=$(0 0 0), and $\bbox{k}_4=$(0 0 1/2). 
The basic magnetic structure, where the magnetic moments on a c-plane lie along the [1 1 0] 
directions with those on the neighboring c-plane along the [1 \={1} 0] directions, 
is described by $\bbox{k}_1$ and $\bbox{k}_2$. 
Canting of the moments from the [1 1 0] directions by an angle of 28$^{\circ}$ 
is described by $\bbox{k}_3$ and $\bbox{k}_4$. 
These properties can naturally be understood by assuming an underlying AFQ order.  
The propagation vector of the AFQ order that is consistent with the basic magnetic structure 
is expected to be (0 0 1/2) if we assume a strong spin-orbit coupling.

The first resonant x-ray scattering experiment on DyB$_2$C$_2$ was performed by 
Hirota {\it et al.}~\cite{Hirota} They discovered two kinds of superlattice reflections of 
$\bbox{k}_2=$(1 0 1/2) and $\bbox{k}_4=$(0 0 1/2) that appear below $T_Q=$25 K. 
From incident energy, temperature, azimuthal angle, and polarization dependences, 
they established that these signals arise from the AFQ order. 
In particular, the characteristic azimuthal-angle dependences for the $\sigma-\sigma'$ 
$(\propto \sin^2 2\varphi)$ and the $\sigma-\pi'$ $(\propto \cos^2 2\varphi)$ 
scatterings were considered to reflect the AFQ order of the $4f$ electrons. 
Another reflection of $\bbox{k}_1=$(1 0 0) was also found to 
appear below $T_{N}=15$ K only for the $\sigma-\pi'$ scattering, reflecting the 
AFM order. 
Tanaka {\it et al.} also performed a similar experiment independently, 
though without polarization analysis, and obtained consistent experimental results.~\cite{Tanaka} 

This paper deals with more comprehensive data and analysis than the first reports of 
Refs.\onlinecite{Hirota} and \onlinecite{Tanaka}. 
The experimental results will be connected quantitatively with the physical picture of the AFQ order 
using the basic formalism described in Sec. \ref{sec:form}. 
Section \ref{sec:exp} describes the experimental procedure. 
We will show in Sec. \ref{sec:result} the experimental results and the analysis 
in the ordered state. We focus 
especially on the explanation of the azimuthal-angle and the polarization dependences 
of the (0 0 5/2) reflection, assuming a model of the AFQ order. 
We also analyze the scatterings of magnetic origin. 
The critical phenomenon of the AFQ ordering is another topic in this paper, which 
is treated in Sec. \ref{sec:crit}. 

\section{Theory}
\label{sec:form}
In order to analyze our experimental results we use the formalism based on symmetry arguments 
developed by Blume.~\cite{Blume2} We summarize the equations in this section. 
Since we deal with the scattering only in the vicinity of the absorpton edge, we do not consider 
the non-resonant terms. The elastic resonant scattering amplitude including up to electric 
quadrupole transition can be written as
\begin{eqnarray}
A_{\mbox{\scriptsize r}}&=& -\frac{e^2}{mc^2}\frac{m\omega_0^{\;3}}{\omega}
   \sum_{\alpha,\beta}\varepsilon_{\beta}'\varepsilon_{\alpha}
   \sum_{\bbox{n},m}e^{i\bbox{\kappa}\cdot (\bbox{n}+\bbox{d}_m)-W_m} 
   \sum_{a,c}p_a \nonumber \\
&& \times
  \sum_{\gamma,\delta}
  \frac{\langle a| R_m^{\beta}-\frac{i}{2}Q_m^{\beta\delta}k_{\delta}' |c\rangle
            \langle c| R_m^{\alpha}+\frac{i}{2}Q_m^{\alpha\gamma}k_{\gamma} |a\rangle }
          {E_a - E_c +\hbar\omega - i\Gamma/2} \label{eq:1} ,
\end{eqnarray}
where 
\begin{eqnarray}
 R_{m}^{\alpha} &=& \sum_{i\in m}r_{i\alpha} \label{eq:2} \\
 Q_{m}^{\alpha\beta} &=& \sum_{i\in m} r_{i\alpha}r_{i\beta} \label{eq:3}
\end{eqnarray}
are the electric dipole and quadrupole moment operators. 
$\bbox{k}(\bbox{k}')$ and $\bbox{\varepsilon}(\bbox{\varepsilon}')$ are 
the wave vector and the polarization vector of the incident(scattered) photon, respectively; 
$\alpha$, $\beta$, $\gamma$, and $\delta$ vary over the cartesian indices $x$, $y$, and $z$. 
The scattering vector is written by $\bbox{\kappa}=\bbox{k}-\bbox{k}'$. 
 The initial and intermediate states of the sample with energies 
$E_a$ and $E_c$ are represented by $|a\rangle$ and $|c\rangle$, respectively; 
$\hbar\omega$ is the energy of the photon and $\hbar\omega_0$ is equal to 
the energy difference $E_c - E_a$. 
$\bbox{n}$ represents the position of the $n$-th unit cell and $\bbox{d}_m$ represents the 
$m$-th atom in the $n$-th unit cell with the Debye-Waller factor $W_m$. In Eqs. (\ref{eq:2}) and 
(\ref{eq:3}) the summations are taken over all the electrons of the $m$-th atom. $p_a$ is the 
thermodynamic probability that the sample is in the state $|a\rangle$. We have introduced 
$\Gamma$, the width of the resonance, which corresponds to the lifetime of the 
intermediate state.

The scattering amplitude for the electric dipole (E1) transition is written as 
\begin{equation}
A_{\mbox{\scriptsize E1}}
= -\frac{e^2}{mc^2} \frac{m\omega_0 ^{\;3}}{\omega}\!
    \sum_{\bbox{n},m}e^{i\bbox{\kappa}\cdot (\bbox{n}+\bbox{d}_m)-W_m}\!
    \sum_{\alpha,\beta}\varepsilon_{\beta}'\varepsilon_{\alpha}f_{m}^{\alpha\beta},
    \label{eq:4}
\end{equation}
where $f_{m}^{\alpha\beta}$ is the atomic scattering factor tensor for the E1 transition 
written by 
\begin{equation}
 f_{m}^{\alpha\beta} = \sum_{a,c} p_a 
  \frac{\langle a| R_m^{\beta} |c\rangle \langle c| R_m^{\alpha} |a\rangle }
          {\hbar\omega -\hbar\omega_0 - i\Gamma/2} \label{eq:5}.
\end{equation}

We assume here that there exists a special axis for the $m$-th atom. The axis is defined, 
for instance, by the magnetic moment, by the quadrupolar moment, or by the local 
crystalline electric field. If we take this axis as the $x$-axis, the atomic scattering factor can be written as 
\begin{eqnarray}
  f&=&d_{0}\left( \begin{array}{ccc} 1&0&0 \\ 0&1&0 \\ 0&0&1 \end{array} \right)
    +id_{1}\left( \begin{array}{ccc} 0&0&0 \\ 0&0&1 \\ 0&-1&0 \end{array} \right) 
    \nonumber \\ 
    && +d_{2}\left( \begin{array}{ccc} 
    \frac{2}{3}&0&0 \\ 0&-\frac{1}{3}&0 \\ 0&0&-\frac{1}{3} \end{array} \right) .
\label{eq:6}
\end{eqnarray}
The parameters $d_{0}$, $d_{1}$, and $d_{2}$ are the coefficients for the 
isotropic, asymmetric, and symmetric parts of the tensor, respectively, 
including the energy dependence 
$m\omega_0^{\;3}/\omega/(\hbar\omega -\hbar\omega_0 - i\Gamma/2)$. 
The $d_1$ term arises purely from the magnetic moment. When the atom 
does not have a magnetic moment, the $d_1$ term vanishes. 
The coefficients $d_0$ and $d_2$ contain both the magnetic and quadrupolar contributions.

With respect to the scattering amplitude for the electric quadrupole (E2) transition, 
we rewrite Blume's equation in a more appealing form. 
We use the quadrupolar operators defined by
\begin{eqnarray}
Q_m^u &=& \sum_{i\in m} (3z_{i}^{\;2}-r_{i}^{\;2}) \nonumber \\
Q_m^v &=& \sum_{i\in m} \sqrt{3}(x_{i}^{\;2} - y_{i}^{\;2}) \nonumber \\
Q_m^{\xi} &=& \sum_{i\in m} 2\sqrt{3}y_i z_i \label{eq:7} \\
Q_m^{\eta} &=& \sum_{i\in m} 2\sqrt{3}z_i x_i \nonumber \\
Q_m^{\zeta} &=& \sum_{i\in m} 2\sqrt{3}x_i y_i \nonumber .
\end{eqnarray}
Then the scattering amplitude for the E2 transition is written as 
\begin{equation}
A_{\mbox{\scriptsize E2}}
= -\frac{e^2}{mc^2} \frac{m\omega_0 ^{\;3}}{4\omega}\!
    \sum_{\bbox{n},m}e^{i\bbox{\kappa}\cdot (\bbox{n}+\bbox{d}_m)-W_m}\!
    \sum_{\alpha,\beta}K_{\beta}'K_{\alpha}g_{m}^{\alpha\beta}  ,
    \label{eq:8}
\end{equation}
where $g_{m}^{\alpha\beta}$ is the atomic scattering factor tensor for the E2 transition 
written by 
\begin{equation}
 g_{m}^{\alpha\beta} = \sum_{a,c} p_a 
  \frac{\langle a| Q_m^{\beta} |c\rangle \langle c| Q_m^{\alpha} |a\rangle }
          {\hbar\omega -\hbar\omega_0 - i\Gamma/2} \label{eq:9}.
\end{equation}
Here the indices $\alpha$ and $\beta$ vary over $u$, $v$, $\xi$, $\eta$, and $\zeta$. 
The coefficient $K$ is calculated to be 
\begin{eqnarray}
K_u &=& \frac{1}{2} \varepsilon_{z}k_{z} \nonumber \\
K_v &=& \frac{1}{2\sqrt{3}}(\varepsilon_{x}k_{x}-\varepsilon_{y}k_{y}) \nonumber \\
K_{\xi} &=& \frac{1}{2\sqrt{3}}(\varepsilon_{y}k_{z}+\varepsilon_{z}k_{y})  \label{eq:10} \\
K_{\eta} &=& \frac{1}{2\sqrt{3}}(\varepsilon_{z}k_{x}+\varepsilon_{x}k_{y}) \nonumber \\
K_{\zeta} &=& \frac{1}{2\sqrt{3}}(\varepsilon_{x}k_{y}+\varepsilon_{y}k_{x}) \nonumber .
\end{eqnarray}

When we assume a special axis for the $m$-th atom and take this axis as the $x$-axis as in 
the case of E1 transition, the atomic scattering factor for the E2 transition 
can be written as 
\begin{eqnarray}
g&=&\left( \begin{array}{ccccc}
             g_{uu}&g_{uv}&0&0&0 \\
             g_{uv}&g_{vv}&0&0&0 \\
             0&0&g_{\xi\xi}&0&0 \\
             0&0&0&g_{\eta\eta}&0 \\
             0&0&0&0&g_{\zeta\zeta} \end{array} \right) \nonumber \\
     &&+
         \left( \begin{array}{ccccc}
           0&0&\sqrt{3}g_{v\xi}&0&0 \\
           0&0&g_{v\xi}&0&0 \\
           -\sqrt{3}g_{v\xi}&-g_{v\xi}&0&0&0 \\
           0&0&0&0&g_{\eta\zeta} \\
           0&0&0&-g_{\eta\zeta}&0  \end{array} \right)   ,
\label{eq:11}
\end{eqnarray}
where
\begin{eqnarray}
g_{uu}&=& 12b_2 + 4c_2 + e_2 - 8f_2 \nonumber \\
g_{uv}&=& -\sqrt{3}(4c_2 + e_2 + 4f_2) \nonumber \\
g_{vv}&=& 12b_2 + 12c_2 +3e_2 \nonumber \\
g_{\xi\xi}&=& 12(b_2 - f_2) \label{eq:12} \\
g_{\eta\eta}&=& 12(b_2 + c_2) \nonumber \\
g_{\zeta\zeta}&=& 12(b_2 + c_2) \nonumber 
\end{eqnarray}
and
\begin{eqnarray}
g_{v\xi}&=& -12a_1 \nonumber \\
g_{\eta\zeta}&=& -12(a_1 + b_1)  \label{eq:13} .
\end{eqnarray}
The coefficients $b_2$, $c_2$, $e_2$, and $f_2$ are for the symmetric part of the 
scattering factor and $a_1$ and $b_1$ are for the asymmetric part, respectively. 
These are the same coefficients used in Ref.~\onlinecite{Blume2} and 
include the same energy dependence as in the E1 transition. 
The asymmetric part is purely of magnetic origin and the symmetric part 
contains both magnetic and quadrupolar contributions. 

We notice from Eq.~(\ref{eq:1}) that there appears the cross term of dipole and quadrupole 
transitions. We do not consider this contribution in this paper because the 
Dy atom is located at the center of symmetry and the cross term vanishes.

The wavevectors and the polarization vectors for the scattering 
configuration illustrated in Fig.~\ref{fg:config} are written as 
\begin{eqnarray}
\bbox{k}&=&k(0, \; \cos\theta, \; -\sin\theta) \nonumber \\
\bbox{k}'&=&k(0, \; \cos\theta, \; \sin\theta) \nonumber \\
\bbox{\varepsilon}_{\sigma}&=&( 1, \; 0, \; 0) \nonumber \\
\bbox{\varepsilon}_{\pi}&=&(0, \; \sin\theta, \; \cos\theta) \label{eq:eps} \\
\bbox{\varepsilon}_{\sigma}'&=&(1, \; 0, \; 0) \nonumber \\
\bbox{\varepsilon}_{\pi}'&=&(0, \; -\sin\theta, \; \cos\theta) \nonumber .
\end{eqnarray}
When the sample is rotated around the $z$-axis by an azimuthal-angle $\varphi$, 
the rotation matrix
\begin{equation}
U(\varphi)=\left (
\begin{array}{ccc}
\cos\varphi & \sin\varphi & 0 \\
-\sin\varphi & \cos\varphi & 0 \\
0 & 0 & 1 
\end{array}
\right)
\label{eq:rot}
\end{equation}
must be operated to all of the vectors in Eq.~(\ref{eq:eps}) from the left.

\section{Experimental}
\label{sec:exp}
The crystal was grown by the Czochralski pulling method with a tetra-arc furnace. 
The obtained single crystal was checked by powder x-ray diffraction, which showed a 
single phase pattern of DyB$_2$C$_2$. The temperature dependence of the magnetic susceptibility 
also agreed with the data reported in Ref.\onlinecite{Yamauchi}. 

X-ray scattering measurements were performed on a four circle diffractometer at BL-16A2 
of the Photon Factory in KEK. A sample with a c-plane surface ($\sim$2$\times$2~mm$^2$) 
was mounted in a closed cycle $^4$He refrigerator so as to align the $c$-axis parallel to the $\phi$-axis 
of the diffractometer, i.e., the $c$-axis is parallel to the $z$-axis in Fig.~\ref{fg:config}. 
The mosaic width was $0.09^{\circ}$ full width at half maximum (FWHM). 
The azimuthal-angle $\varphi$ is defined to be zero when the $b$-axis is parallel to 
the scattering plane. 
The azimuthal-angle scan can be performed by rotating the 
$\phi$-axis of the diffractometer. 

The incident x-ray is almost linearly polarized with its electric field perpendicular to the scattering 
plane ($\sigma$ polarization).  
The polarization of the diffracted beam, i.e., $\sigma'$ (perpendicular to the scattering plane) or $\pi'$ 
(parallel to the scattering plane), was analyzed using the PG(0 0 6) reflection. The scattering 
angle of this reflection is about 91$^{\circ}$ around the $L_3$ edge of Dy, 
resulting in almost perfect analysis. The contamination of the $\pi$ component in the incident 
x-ray was estimated to be 1.5\% from the intensity ratio $\pi-\pi'/\sigma-\sigma'$ of the 
(0 0 2) fundamental reflection. 

\section{Ordered state}
\label{sec:result}
\subsection{experimental results}
The investigation of the reflection corresponding to $\bbox{k}_4$=(0 0 1/2) was performed 
using the (0 0 5/2) reflection. The incident energy dependence of the intensity at 
30~K, 20~K and 10~K, which correspond to the phases I, II, and III, respectively, 
is shown in Fig.~\ref{fg:Edepk4}. The measured peak top intensities have been transformed 
into the integrated intensities by multiplying the width so as to be compared with other figures 
in the same scale. To compare the integrated intensities with the calculated intensities, 
the data have been corrected for the absorption and for the Lorentz factor; 
the absorption coefficient was deduced from the fluorescence spectrum.~\cite{Hirota,Dumesnil} 

Resonant peaks are clearly observed in the spectra at 10~K and 20~K, while 
they are not observed at 30~K. 
The main-edge peak at 7.792 keV 
corresponds to the resonance due to the $2p\leftrightarrow 5d$ electric dipole transition. 
The pre-edge peak at 7.782 keV that is well resolved in the $\sigma-\pi'$ process is probably 
attributed to the $2p\leftrightarrow 4f$ electric quadrupole transition. These assignments 
are consistent with the previous experiments on substances including Dy or other rare-earth 
elements.~\cite{Dumesnil,Hannon,Lang,Bartolome,Isaacs,Gibbs1,Gibbs2} 
Concerning the $\sigma-\sigma'$ process it was not possible to decide if there was any resonance 
at 7.782 keV because of the wide peak at the main-edge. 
A very small peak was observed at 10 K for the $\sigma-\sigma'$ 
process at $\varphi=0^{\circ}$. 
This result is rather confusing since the peak is located between 7.792 keV and 7.782 keV. 

The temperature dependences of the integrated intensities have been measured for each 
resonant peak in Fig.~\ref{fg:Edepk4}.~\cite{Hirota} 
The resonant peaks which are observed both at 10 K and 20 K appear below $T_{Q}=25$ K 
(Fig.~\ref{fg:Edepk4} (b) and (c)), while those which are observed only at 10 K 
appear below $T_{N}=$15 K (Fig.~\ref{fg:Edepk4} (a) and (d)).  
Then the former peaks can be attributed to quadrupolar origin and the latter to magnetic origin. 
With regard to the main-edge and pre-edge peaks in Fig.~\ref{fg:Edepk4} (c), 
the intensities at $\varphi=0^{\circ}$ exhibit the same temperature dependence. 

Figure~\ref{fg:Azik4} shows the azimuthal-angle dependences 
of the integrated intensity at 7.792 keV. 
Peak profile was measured by the $\theta-2\theta$ scan for each point and was fit to a Gaussian. 
The intensity of the $\sigma-\sigma'$ process exhibits $\sin^{2}2\varphi$ dependence 
below $T_{Q}$ . The intensity of the $\sigma-\pi'$ process exhibits $\cos^{2}2\varphi$ 
dependence at temperatures between $T_{Q}$ and $T_{N}$, 
and some magnetic contribution is added below $T_{N}$. 
The azimuthal-angle dependence of the pre-edge peak for the $\sigma-\pi'$ process 
is shown in Fig.~\ref{fg:Azik4pre}. The intensity exhibits completely the same 
angle dependence with that of the main-edge, i.e., $\propto \cos^2 2\varphi$. 
With regard to the $\sigma-\sigma'$ process, the intensity at 7.782 keV also exhibits 
the same azimuthal-angle dependence with that at 7.792 keV. However, it was not 
possible to separate the pre-edge component from the tail of the main-edge peak. 

The incident energy dependences of the integrated intensity of the 
(1 0 2) and (1 0 5/2) reflections, which correspond to $\bbox{k}_1$ and $\bbox{k}_2$, 
respectively, are shown in Fig.~\ref{fg:Edepk2}. 
The scale of the vertical axis is the same as those of Figs.~\ref{fg:Edepk4}--\ref{fg:Azik4pre}.
The measurement of the temperature dependence of these resonant peaks 
shows that they appear below $T_{N}$, indicating magnetic origin. 
It should be noted that there is a small shoulder at the lower energy side 
of the main-edge, which is more clearly observed in the (1 0 5/2) reflection. 
This is considered to be of electric quadrupole transition. 
With regard to the $\sigma-\sigma'$ process, no signal was observed at (1 0 2). 
On the other hand, the (1 0 5/2) reflection exhibits non-resonant type energy 
dependence which appear below $T_{Q}$.~\cite{Hirota} This indicates that 
a periodic lattice distortion occurs simultaneously with the AFQ ordering. This
point will be discussed in Sec. \ref{sec:discuss}.

\subsection{analysis}
\subsubsection{model calculation}
Let us analyze the above experimental results using the formalism described in Sec.~\ref{sec:form}. 
Figure~\ref{fg:AFQmodel} illustrates a model of the AFQ order 
of the $4f$ electrons of the Dy ions, which is expected from the magnetic structure in the phase III. 
We introduce the canting angle $\alpha$ 
as a parameter. 
We assume each ion has its own special axis depending on the local quadrupolar 
or magnetic moment, around which the charge distribution is symmetric. 
We define the direction of the magnetic moment in the phase III 
as the $x$-axis. Since the spin-orbit coupling is strong, the $x$-axis and the principal 
axis of the quadrupolar moment coincides. 
To calculate the scattering amplitudes $A_{\mbox{\scriptsize E1}}$ and 
$A_{\mbox{\scriptsize E2}}$ at an azimuthal-angle $\varphi$, 
it is necessary to rotate the vectors in Eq.~(\ref{eq:eps}) so that the $xyz$-axes in Fig.~\ref{fg:config} 
coincide with those in Fig.~\ref{fg:AFQmodel} for each ion, i.e., the rotation of 
$-\pi/4+\alpha+\varphi$ for Dy(1), $3\pi/4-\alpha+\varphi$ for Dy(2), $\pi/4+\alpha+\varphi$ for Dy(3), 
and $-3\pi/4-\alpha+\varphi$ for Dy(4) is performed using Eq.~(\ref{eq:rot}). 

The resonant scattering cross-section is equal to the square of the scattering 
amplitude $A_{\mbox{\scriptsize r}}$. 
The intensities of the (0 0 5/2) resonant scattering for the electric dipole (E1) transition 
are calculated to be 
\begin{eqnarray}
     |A_{\bbox{k}_4,\mbox{\scriptsize E1}}^{\sigma-\sigma'}|^2 & \propto & 
       |2d_2 \cos 2\alpha \sin 2\varphi|^2 \label{eq:k4e1ss}\\
     |A_{\bbox{k}_4,\mbox{\scriptsize E1}}^{\sigma-\pi'}|^2 & \propto & 
       |2d_2\cos 2\alpha \cos 2\varphi \sin\theta \nonumber \\
       &&+ 2\sqrt{2}id_1 \sin \alpha \sin\varphi \cos\theta|^2 \label{eq:k4e1sp},
\end{eqnarray}
and for the electric quadrupole (E2) transition,
\begin{eqnarray}
     |A_{\bbox{k}_4,\mbox{\scriptsize E2}}^{\sigma-\sigma'}|^2 & \propto & 
       |2(c_2+f_2)\cos 2\alpha \sin 2\varphi \sin^2\theta \nonumber \\
        &&+ 2\sqrt{2}\{a_1+b_1 (2\cos 2\alpha \cos 2\varphi + \cos 2\varphi  \nonumber \\
        && - \cos 2\alpha)\} \cos\varphi \sin \alpha \sin 2\theta|^2 /16 \label{eq:k4e2ss}\\
     |A_{\bbox{k}_4,\mbox{\scriptsize E2}}^{\sigma-\pi'}|^2 & \propto & 
       |\frac{1}{2}\{(4c_2+e_2+4f_2) +(4c_2+e_2+4f_2 \nonumber \\
       && +4c_2+4f_2)\cos 2\theta\} \cos 2\alpha \cos 2\varphi \sin\theta \nonumber \\
        &&+ (\mbox{terms of $a_1$ and $b_1$}) |^2 /16 \label{eq:k4e2sp}.
\end{eqnarray}

Secondly, the intensities of the (1 0 2) resonant scattering at $\varphi=0^{\circ}$ 
for the E1 transition are calculated to be 
\begin{eqnarray}
  |A_{\bbox{k}_1,\mbox{\scriptsize E1}}^{\sigma-\sigma'}|^2 & \propto & 0 
      \label{eq:k1e1ss} \\
  |A_{\bbox{k}_1,\mbox{\scriptsize E1}}^{\sigma-\pi'}|^2 & \propto & 
     |0.63\sqrt{2}id_1 \sin\theta \cos\alpha|^2 \label{eq:k1e1sp} , 
 \end{eqnarray}
 and for the E2 transition 
 \begin{eqnarray}
  |A_{\bbox{k}_1,\mbox{\scriptsize E2}}^{\sigma-\sigma'}|^2 & \propto &
   |(2.7a_1\cos 2\alpha + 0.94b_1 \cos 2\alpha \nonumber \\
    && - 1.2b_1 \cos 3\alpha)\sin 2\theta|^2 /16  \label{eq:k1e2ss} \\
  |A_{\bbox{k}_1,\mbox{\scriptsize E2}}^{\sigma-\pi'}|^2 & \propto &
   |(\mbox{terms of $a_1$, $b_1$, and $e_2$} ) |^2 /16  \label{eq:k1e2sp} .
\end{eqnarray}

Finally, the scattering intensities of the (1 0 5/2) resonant scattering at $\varphi=0^{\circ}$ 
for the E1 transition are calculated to be 
\begin{eqnarray}
  |A_{\bbox{k}_2,\mbox{\scriptsize E1}}^{\sigma-\sigma'}|^2 & \propto & 
    |1.9d_2 \sin 2\alpha|^2  \label{eq:k2e1ss} \\
  |A_{\bbox{k}_2,\mbox{\scriptsize E1}}^{\sigma-\pi'}|^2 & \propto & 
    |2\sqrt{2}id_1\cos\alpha \cos\theta+0.5d_2\sin 2\alpha \cos\theta|^2 
       \label{eq:k2e1sp},
\end{eqnarray}
and for the E2 transition
\begin{eqnarray}
  |A_{\bbox{k}_2,\mbox{\scriptsize E2}}^{\sigma-\sigma'}|^2 & \propto & 
    |2\{0.47(c_2+f_2)+0.031e_2\}\sin 2\alpha \cos 2\theta \nonumber \\
    && -\{1.1(c_2+f_2)+0.061e_2\}\sin 2\alpha |^2 /16 \label{eq:k2e2ss} \\
  |A_{\bbox{k}_2,\mbox{\scriptsize E2}}^{\sigma-\pi'}|^2 & \propto & 
    |(\mbox{terms of $a_1$, $b_1$, $e_2$, and $c_2 + f_2$} ) |^2 /16 \label{eq:k2e2sp} .
\end{eqnarray}

The factor of $|e^2/mc^2|^2$ 
has been omitted in the above expressions to save space. 
The expressions for the $\sigma-\pi'$ process for the E2 
transition are not written explicitly since there appear many terms; 
only the parameters which appear are written. 

\subsubsection{dipole transition}
The azimuthal-angle dependences of the (0 0 5/2) reflection at 7.792 keV in 
Fig.~\ref{fg:Azik4} are well reproduced by Eqs.~(\ref{eq:k4e1ss}) 
and (\ref{eq:k4e1sp}). The $\sin^2 2\varphi$ dependence of the $\sigma-\sigma'$ 
process and the $\cos^2 2\varphi$ dependence of the $\sigma-\pi'$ process are 
explained by the $d_2$ term. 
Between $T_{N}$ and $T_{Q}$, $d_1$ vanishes and $d_2$ arises only from the 
quadrupolar moment of the $4f$-electrons. The intensity ratio 
$I^{\sigma-\pi'}/I^{\sigma-\sigma'}\approx 0.3$ can be explained by 
$\sin^2 \theta=0.3155$ for the Bragg angle of the (0 0 5/2) reflection. 
The $d_1$ term contributes below $T_{N}$. 
If we assume $\alpha=28^{\circ}$ as determined in Ref.~\onlinecite{Yamauchi}, 
we obtain $d_1=(f_{yz}-f_{zy})/2i=0.91$ and $d_2=(f_{xx}-f_{yy})=2.6$ 
to explain quantitatively the integrated intensities of 
the (0 0 5/2) reflection at 10 K by Eqs.~(\ref{eq:k4e1ss}) and (\ref{eq:k4e1sp}). 
By comparing the integrated intensity with that of the (0 0 2) fundamental reflection 
(2$\times 10^5$), which is ascribed to the Thomson scattering from $4\times 88$ electrons, 
we can deduce that the value of $d_2=2.6$ corresponds to 1.4 electrons 
(per four molecules). 

Below $T_N$ the ordered magnetic moment also contributes to $d_2$ through 
an exchange splitting and a spin polarization of the $5d$ level.~\cite{Hannon} 
This should manifest below $T_N$ in the temperature dependence of the intensity. 
However, the intensity for the $\sigma-\sigma'$ process at $\varphi=45^{\circ}$ does 
not show any clear kink in the temperature dependence around $T_N$. 
This result indicates that $d_2$ is caused mostly by the quadrupolar moment 
through Coulomb interaction between $4f$ and $5d$ electrons. 
The contribution of the magnetic moment, which is estimated to be $7.1\mu_{B}$ 
in Ref.~\onlinecite{Yamauchi}, seems much smaller than that of the quadrupolar moment. 

Using Eqs.~(\ref{eq:k1e1sp}) and (\ref{eq:k2e1sp}), 
the above parameters at 10 K, i.e., $\alpha=28^{\circ}$, $d_1=0.91$, and $d_2=2.6$, 
give the integrated intensities of 0.11 for the (1 0 2) reflection and 4.17 for the 
(1 0 5/2) reflection, respectively. The value of 0.11 for the (1 0 2) reflection does not 
reproduce the experimental result of about 1.5 in Fig.~\ref{fg:Edepk2}. 
This can be ascribed to the experimental difficulty of rotating the crystal to different 
reciprocal lattice points without changing the effective volume which contributes 
to the scattering. However, the calculation at least explains these two magnetic peaks 
qualitatively. 

According to Eq. (\ref{eq:k4e1sp}) the intensity of the (0 0 5/2) reflection 
for the $\sigma-\pi'$ process at $\varphi=45^{\circ}$ is proportional 
to $|d_1\sin \alpha|^2$. The fact that a finite intensity is observed at 10 K 
indicates that the canting angle of the magnetic moment is certainly not zero 
below $T_N$. 
On the other hand, Eq. (\ref{eq:k2e1sp}) shows that the intensity of the 
(1 0 5/2) reflection for the $\sigma-\pi'$ scattering at $\varphi=0^{\circ}$ 
is proportional to $|d_2\sin 2\alpha|^2$ above $T_N$ since the $d_1$ term vanishes. 
The experimental result in Fig.~\ref{fg:Edepk2} shows that the intensity 
completely disappears at 20 K. This indicates that the canting angle 
of the quadrupolar moment could be zero at 20 K since $d_2$ is definitely not zero 
at this temperature.
Then, it is suggested that the canting of the moments occurs only below $T_N$. 

From Eq.~(\ref{eq:k2e1ss}) we see that 
the $\sigma-\sigma'$ scattering of the (1 0 5/2) reflection can also be a measure 
of the canting angle. However, since this scattering shows a non-resonant type energy 
dependence as a result of a periodic lattice distortion,~\cite{Hirota} it was difficult 
to extract the resonant contribution at 7.792 keV with sufficient accuracy to 
examine the behavior of the canting angle in detail. 

\subsubsection{quadrupole transition}
The calculated results are qualitatively consistent with the experimental results. 
The azimuthal-angle dependence of the (0 0 5/2) reflection for the 
$\sigma-\pi'$ process at 7.782 keV shown in 
Fig.~\ref{fg:Azik4pre} is well reproduced by the quadrupolar terms of 
$4c_2+e_2+4f_2=-g_{uv}/\sqrt{3}$ and $4(c_2+f_2)=-(g_{\xi\xi}-g_{\eta\eta})/3$ in 
Eq.~(\ref{eq:k4e2sp}). The quadrupolar term of Eq.~(\ref{eq:k4e2ss}) 
is also consistent with the experiment. 
The very small intensities in Fig.~\ref{fg:Edepk4} (a) and (d) might be the 
magnetic signal from $a_1$ and $b_1$ terms. 

The magnitudes of $g_{uv}$ and $(g_{\xi\xi}-g_{\eta\eta})$ 
could be estimated by comparing the intensity of the $\sigma-\pi'$ process 
with that of the $\sigma-\sigma'$ process. 
However, we were not able to estimate the pre-edge peak intensity for the 
$\sigma-\sigma'$ process because of the difficulty in resolving the resonant peak into 
two peaks of E1 and E2 transitions. 

With regard to the $\sigma-\pi'$ scattering of the (1 0 2) and (1 0 5/2) reflections, 
the small shoulders around the pre-edge in Fig.~\ref{fg:Edepk2} can be ascribed to the 
magnetic signal due to the $a_1$ and $b_1$ terms in Eqs.~(\ref{eq:k1e2sp}) and (\ref{eq:k2e2sp}), 
respectively. The quadrupolar terms of $e_2$ and $c_2+f_2$ have the factor of $\sin 2\alpha$ 
or $\sin 4\alpha$. 
Then, the fact that the intensity disappears above $T_N$ 
is consistent with the argument on the canting angle in the previous subsection. 

\section{critical phenomenon}
\label{sec:crit}
Our interest here is to investigate the critical phenomena associated with 
the AFQ ordering by measuring the peak profiles precisely, especially around the 
transition temperature $T_{Q}$.  While almost all the orbital orderings in manganese 
oxides exhibit first-order-like phase transitions, the AFQ ordering in DyB$_2$C$_2$ 
is considered to be of second order. 
The information on the critical phenomena may give some insight into the mechanism of the 
interaction between the quadrupole moments.

We have utilized a Ge (111) crystal as an 
analyzer to obtain good resolution to observe diffuse scatterings. 
The longitudinal (0 0 $l$) scans for both the (0 0 2) fundamental Bragg peak and the well 
developed (0 0 5/2) superlattice Bragg peak at low temperature were able to be reproduced 
by a squared Lorentzian with its FWHM 0.0011 \AA$^{-1}$. 
We have therefore assumed this as the resolution function for the present measurement. 

We have concentrated in this study on the (0 0 5/2) reflection at $\varphi=45^{\circ}$. 
Although the polarization analysis was not performed, we already know from 
Fig.~\ref{fg:Azik4} that this reflection consists only of the $\sigma-\sigma'$ scattering 
above $T_{N}$. The measurement was performed much more precisely than the one in Ref.\onlinecite{Hirota}. 
Although the temperature stability of the thermometer was kept within $\pm 0.01$ K, heating of the sample 
by the beam caused a serious problem in the measurement of the temperature dependence 
of the weak signals at the critical region. The problem is that the beam intensity changes with the ring current, 
which decreases typically from 400 mA to 250 mA in a day. This leads to different heating powers. 
It reached about 0.25~K at most when the beam intensity was high. Therefore, very careful data taking 
and treatment of the data, namely shifting of the temperature, were necessary. 

The integrated intensities and the peak widths were obtained by fitting the profiles to a squared Lorentzian. 
Figure~\ref{fg:diffuse1} shows the obtained temperature dependence of the integrated intensity 
of the (0 0 5/2) reflection for the longitudinal (0 0 $l$) scans. 
Note that the scale of the vertical axis is different from the other figures. 
The intensities below $T_{Q}$ were fitted 
to a power law $I\propto ((T_{Q}-T)/T_{Q})^{2\beta}$ with varying the fitting range 
from 15 K -- 25.5 K to 23 K -- 25.5 K. 
The obtained parameters are $T_{Q}=25.52\pm 0.009$ K and $\beta=0.35\pm 0.01$, which 
are demonstrated by the solid line in the figure. 
This $\beta$ is different from the previously reported value 
of 0.18 in Ref.\onlinecite{Hirota}; this is probably because the previous measurement 
did not have enough accuracy to determine the critical exponent due to small number of 
data points. 

The inset in Fig.~\ref{fg:diffuse1} shows the integrated intensity around $T_{Q}$. Although a careful 
data treatment was necessary to estimate the reliable temperatures as described above, 
it is certain that the intensity does not vanish even above $T_{Q}$. Increase in the peak width was also 
observed around $T_{Q}$. 
The FWHM of the original profile was deconvoluted with the resolution function and was 
converted to the inverse correlation length along the $c$-axis. 
The result is shown in Fig.~\ref{fg:diffuse2}. Although $\kappa_c$ increases with increasing the 
temperature above $T_{Q}$, it was not possible to deduce the critical exponent from these small 
number of points.

\section{Discussions}
\label{sec:discuss}
\subsection{four propagation vectors} 
We can interpret the appearance of the four propagation vectors 
in the following way. At $T_Q=25$ K 
the AFQ order with $\bbox{k}_4=$(0 0 1/2) occurs. The AFQ moment is 
the principal order parameter and induces the periodic lattice distortion 
with $\bbox{k}_2=$(1 0 1/2) through some quadrupole-strain couplings. 
It should be noted that (1 0 1/2) and (0 0 1/2) are the equivalent reciprocal lattice 
points. The resonance can occur also at (1 0 1/2) simultaneously with the resonance at (0 0 1/2). 
However, we could not identify the resonance at (1 0 5/2) 
because the non-resonant scattering was dominant in the $\sigma-\sigma'$ process 
or because the canting angle of the quadrupolar moment could be zero in the phase II. 

Below $T_N=15$ K the AFM order with $\bbox{k}_1=$(1 0 0) occurs. 
The AFM moment becomes the additional order parameter in this phase. 
Since (1 0 0) and (0 0 0) are the equivalent reciprocal lattice points, 
the magnetic Bragg peak also appears at $\bbox{k}_3=$(0 0 0). 
Furthermore, since there is already an AFQ order which confines the 
direction of the magnetic moments, the resultant magnetic structure also 
gives magnetic Bragg peaks at (1 0 1/2) and at (0 0 1/2). 
Thus, the four propagation vectors are coupled with each other. 

\subsection{periodic displacement of atoms}
The non-resonant scattering of the (1 0 1/2) reflection that appears below 
$T_{Q}$ indicates a periodic displacement of the atoms, 
which leads to the Bragg reflection due to the Thomson scattering. 
Tanaka {\it et al.} interpreted this reflection as a displacement of B and C atoms 
which reduces the symmetry to the space group $P4_2/mnm$.~\cite{Tanaka} 
Lovesey and Knight recently gave a theoretical calculation,~\cite{Lovesey} 
in which they explained the (0 0 1/2) resonant peak from the same standpoint 
as in Ref.~\onlinecite{Tanaka}. 
However, we consider that the displacement of B and C atoms is not realistic 
because the estimated scattering factors of the possible 
reflections from such structure are not consistent with the experimental 
results. If we assume the $z$ parameter of the $8j$ site of $P4_2/mnm$, where the 
B and C atoms are located, shifted from the original value of 0.5 to 0.45, 
the squared structure factor for the (1 0 5/2) reflection becomes only 0.0376 
whereas those for (1 1 3/2) and (2 2 3/2) becomes 154.6 and 170.6, respectively. 
Therefore, much stronger reflections are expected for the two points. However, 
none of these reflections have been observed. Furthermore, neutron powder-diffraction 
experiments also do not show any evidence of the displacement of B and C atoms.~\cite{Yamauchi2} 

Since the atomic scattering factor of a Dy ion is much larger than that of B and C, 
it is natural to attribute the non-resonant (1 0 1/2) reflection to a periodic displacement  
of the Dy ions. 
We propose a model, where the Dy(1) and Dy(4) in Fig.~\ref{fg:AFQmodel} are displaced by 
$+\delta$ along the $c$-axis and the Dy(2) and Dy(3) by $-\delta$. This gives the 
superlattice reflection at (1 0 1/2). 
The structure factor for the (1 0 5/2) reflection is calculated to be 
$F=-4if_{{\mbox{\scriptsize Dy}}}\sin 5\pi \delta /c$, while that for the (0 0 2) fundamental reflection 
is $8f_{{\mbox{\scriptsize B}}}+8f_{{\mbox{\scriptsize C}}}+
4f_{{\mbox{\scriptsize Dy}}}\cos 4\pi\delta /c$. From the observed intensity ratio 
between (1 0 5/2) and (0 0 2), 
a reasonable value of $\delta /c \approx 4\times 10^{-4}$ is deduced. 

We consider this displacement to be related with the characteristic crystal structure, where 
the Dy ions are located between the hard B-C layers. This displacement corresponds 
to one of the 30 phonon modes of DyB$_2$C$_2$ at $\bbox{k}=$(0 0 1/2). 
Though there are many other phonon modes that involve 
displacements of B and C, including the one proposed in Ref.~\onlinecite{Tanaka}, 
their energy must be very high because they have to modify 
the strong covalent bonding among the B-C network. On the contrary, 
the energy scale of the motion of the Dy ions is expected to be small. Then, the position 
of the Dy ions is considered to be susceptible to quadrupolar orderings through 
a coupling with the lattice. 

There is another reason that the resonant scattering in DyB$_2$C$_2$ reflects 
the AFQ order itself but not the symmetry of the crystal. 
The local symmetries of Dy(1) and Dy(2), as numbered 
in Fig.~\ref{fg:AFQmodel}, are different even in the original crystal structure above $T_Q$. 
This can give the (1 0 0) resonant scattering even above $T_Q$, which is not 
observed experimentally. Then, even though there was any displacement 
of B and C atoms, the effect on the resonant scattering would be 
negligible. In addition, the displacement of the Dy ions as we propose 
is mere $4\times 10^{-4}$, also negligibly small to give any effect on the resonant scattering. 
Finally, it should be noted that the theory in Ref.~\onlinecite{Lovesey} can also be applied to 
our model of the AFQ order without the displacement of B and C but with the same 
space group $P4_2/mnm$; since the theory is based only on the local 
symmetry of Dy, it is not restricted by the mechanism of the symmetry lowering. 

\subsection{critical phenomenon}
It has been clarified from the present experiment that the order parameter vanishes continuously 
through $T_{Q}$ and that there is still nonvanishing intensity even above $T_{Q}$. 
The temperature dependence of the order parameter below $T_{Q}$ well follows a 
normal power law with a critical exponent $\beta=0.35\pm 0.01$, which is not an anomalous value 
when compared with the 3D-Heisenberg model (0.365),~\cite{Guillou} 
3D-XY model (0.345),~\cite{Guillou} and actual 
magnetic systems such as EuO (0.36)~\cite{Nielsen} and MnF$_2$ (0.31).~\cite{Goldman} 
These results indicate that the AFQ ordering in DyB$_2$C$_2$ is really a second-order 
phase transition. 
Note that we assume in this paper that the intensity of the resonant scattering is proportional 
to the square of the order parameter. Ishihara and Maekawa discuss this relation 
in Ref.\onlinecite{Ishihara3}.

Concerning the diffuse scattering, it was very difficult to measure precisely the broadening 
of the width with increasing temperature due to the weak intensity in comparison with the 
background. However, it should be noted that this result does not directly mean the absence of diffuse 
scattering. The present situation seems very similar to the case of critical magnetic scattering 
in MnF$_2$ studied by x-ray scattering.~\cite{Goldman} It was also not possible to measure diffuse 
scattering above $T_{N}=67.4$ K due to weak count rates. However, the diffuse scattering 
certainly exists and were measured up to 10 K above $T_{N}$ by neutron scattering.~\cite{Schulhof} 
The width of the neutron scattering profile was ten times larger than that of the x-ray 
scattering.~\cite{Schulhof} 

This problem might be related with the so called two length scales problem. 
There are some systems in which peak profiles of diffuse scatterings above transition temperatures 
consist of a narrow central peak and a broad one. Detailed studies on this problem using both 
neutrons and x-rays showed that x-ray signals are dominated by narrow 
components because of its small resolution volume, while broad component manifests in 
neutron scattering.~\cite{Thurston,Hirota_2} 
Though the origin of the narrow component has not yet been established, it is interpreted 
as the near surface effect that is more sensitive for x-rays.~\cite{Thurston,Hirota_2} 

The observed inverse correlation length $\kappa_c$ of DyB$_2$C$_2$ above $T_{Q}$ is about 
$5\times 10^{-4}$ \AA$^{-1}$ at a reduced temperature $(T-T_{Q})/T_{Q}=4\times 10^{-3}$, 
which is read from Fig.~\ref{fg:diffuse2}. This value is as large as that of the 
narrow component of holmium measured by resonant x-ray scattering at the same reduced temperature 
$(T-T_{N})/T_{N}$.~\cite{Thurston} Furthermore, in holmium, broad diffuse scattering 
with its width ten times wider than the narrow component of x-ray is certainly observed by 
neutron scattering.~\cite{Thurston} 
Therefore, we should not conclude that there is no broad diffuse scattering in  DyB$_2$C$_2$ 
only from the present x-ray scattering study.

\section{Conclusions}
We have performed the resonant x-ray scattering on DyB$_2$C$_2$ 
and have investigated the signals that correspond to the AFQ and AFM orders. 
The experimental results are analyzed both quantitatively and qualitatively using a theory 
of resonant x-ray scattering and a model of the AFQ order. 
An important result is that the quadrupolar order parameter manifests 
especially in the $\sigma-\sigma'$ scattering which little contains the 
magnetic contribution. We analyzed the resonant peak at (0 0 1/2) which corresponds 
to the AFQ order using a parameter $d_2=(f_{xx}-f_{yy})$ in this paper. 

The (1 0 1/2) reflection is also an important peak for more detailed study of the AFQ order 
in this compound. One reason is that the 
non-resonant $\sigma-\sigma'$ scattering suggests a periodic displacement of the Dy 
ions below $T_Q$, which is probably caused by a quadrupole-strain coupling that is 
peculiar in this compound. Another reason is that a detailed investigation of the 
resonant peak could reveal the behavior of the canting angle of the moments 
especially in the phase II, i.e., if it is zero or not. 
Both of these subjects require further studies. 

We have also studied the critical phenomenon of the AFQ ordering. The order parameter well 
follows a normal power law with a reasonable critical exponent for a 3D system. The 
second-order nature of the phase transition was confirmed from the continuous decrease of the 
order parameter and the diffuse scattering above $T_{Q}$. 

Finally, although the intra-atomic $d-f$ Coulomb interaction is expected to be the most 
probable origin of the anisotropic tensor of x-ray susceptibility in the $4f$-electron systems, 
we need further investigation, both theoretically and experimentally, to clarify the mechanism of the 
scattering process.

\acknowledgments
We are deeply indebted to H. Yamauchi, K. Indoh, K. Ohoyama, H. Onodera, 
and Y. Yamaguchi for useful information about sample preparation and 
fruitful discussions. We also acknowledge N. Kimura for collaboration in 
growing single crystals.
This work was partly supported by a Grant-in-Aid for Scientific Research 
from the Ministry of Education, Science, Sports and Culture, by Core 
Research for Evolutional Science and Technology (CREST) from Japan Science 
and Technology Corporation, and by REIMEI kenkyu from Japan Atomic Energy 
Research Institute.

\appendix

% now the references. delete or change fake bibitem. delete next three
%   lines and directly read in your .bbl file if you use bibtex.00.7.11


\begin{figure}
\caption{
(left) Crystal structure of DyB$_2$C$_2$ ($P4/mbm$, $a$=5.341 \AA, $c$=3.547 \AA\ at 30 K). 
The magnetic structure is indicated by the arrows. 
(right) The $h$-$l$ plane of the reciprocal space. Black marks are the reflection points 
that were actually investigated in the present experiment. }
\label{fg:structure}
\end{figure}

\begin{figure}
\caption{The definition of the vectors associated with the x-rays and the 
axes attatched to the crystal.}
\label{fg:config}
\end{figure}

 \begin{figure}
 \caption{
Incident energy dependences of the integrated intensity of the (0 0 5/2) reflections 
corrected for the absorption and the Lorentz factor: 
(a) $\sigma-\sigma'$ scattering at $\varphi=0^{\circ}$, 
(b) $\sigma-\sigma'$ scattering at $\varphi=45^{\circ}$, 
(c) $\sigma-\pi'$ scattering at $\varphi=0^{\circ}$, and
(d) $\sigma-\pi'$ scattering at $\varphi=45^{\circ}$. 
Note that the integrated intensity of the (0 0 2) fundamental peak is 2$\times 10^5$. 
}
 \label{fg:Edepk4}
 \end{figure}

\begin{figure}
\caption{Azimuthal-angle dependences of the integrated intensity of the (0 0 5/2) 
reflection for the $\sigma-\sigma'$ and the $\sigma-\pi'$ scatterings 
at the main-edge. 
Solid lines are the fits with $\sin^2 2\varphi$ for $\sigma-\sigma'$ and 
with $\cos^2 2\varphi$ for $\sigma-\pi'$.}
\label{fg:Azik4}
\end{figure}

\begin{figure}
\caption{Azimuthal-angle dependence of the integrated intensity of the (0 0 5/2) reflection 
for the $\sigma-\pi'$ scattering at the pre-edge. Solid line is a fit with $\cos^2 2\varphi$. }
\label{fg:Azik4pre}
\end{figure}

\begin{figure}
\caption{Incident energy dependences of the integrated intensity for the 
$\sigma-\pi'$ scatterings at $\varphi=0^{\circ}$ 
corrected for the absorption and the Lorentz factor: 
(a) (1 0 2) reflection and 
(b) (1 0 5/2) reflection. 
}
\label{fg:Edepk2}
\end{figure}

\begin{figure}
\caption{A model of the antiferroquadrupolar order in DyB$_2$C$_2$. 
The shadows represent the anisotropic charge distributions. 
The unit cell is expressed by a$\times$a$\times$2c, which contains four Dy ions. 
The canting angle $\alpha$ from the [1 1 0]-equivalent axes is treated as a parameter. 
The direction of the magnetic moment in the phase III is taken as the $x$-axis. 
}
\label{fg:AFQmodel}
\end{figure}

\begin{figure}
\caption{Temperature dependence of the integrated intensity of the (0 0 5/2) reflection 
for the (0 0 $l$)-scan at $\varphi=45^{\circ}$ measured with a Ge(111) analyser. 
Solid line is a fit to a power law $I\propto ((T_{Q}-T)/T_{Q})^{2\beta}$. 
Inset shows the integrated intensity around the transition temperature. }
\label{fg:diffuse1}
\end{figure}

\begin{figure}
\caption{Temperature dependence of the inverse correlation length along the $c$-axis 
obtained by deconvoluting the peak widths to the resolution width. Solid line is a 
calculated curve for the critical exponent $\nu=0.7$. Inset shows the peak profile 
at $T=25.6$ K. Solid line is a fit to a squared Lorentzian and dotted line is a 
squared Lorentzian with the same height and with the resolution width.}
\label{fg:diffuse2}
\end{figure}

\end{document}